\documentclass[]{aastex6}
\usepackage{graphicx,epsfig,float,rotate,amsmath,amssymb,natbib}
\pdfoutput=1

\slugcomment{Draft, \today, To be submitted to ApJ}
\shorttitle{Zombie Vortex Instability II}
\shortauthors{MPJB2016}

\newcommand*{\eg}{e.g.,\ }
\newcommand*{\ie}{i.e.,\ }

\numberwithin{equation}{section}

\defcitealias{MPJH13}{MPJH13}
\defcitealias{MPJBHL2015}{MPJBHL15}

\begin{document}

\title{ZOMBIE VORTEX INSTABILITY. II.  THRESHOLDS TO TRIGGER INSTABILITY AND THE PROPERTIES OF ZOMBIE TURBULENCE IN THE DEAD ZONES OF PROTOPLANETARY DISKS}

\author{Philip S. Marcus, Suyang Pei\altaffilmark{1}, Chung-Hsiang Jiang}
\affil{Department of Mechanical Engineering, University of California, Berkeley\\ 6121 Etcheverry Hall, Mailstop 1740, Berkeley, CA 94720-1740}
\email{pmarcus@me.berkeley.edu}
\altaffiltext{1}{Current affiliation: Department of Physical \& Environmental Sciences, Texas A\&M University, Corpus Christi}

\author{Joseph A. Barranco}
\affil{Department of Physics \& Astronomy, San Francisco State University\\ 1600 Holloway Avenue, San Francisco, CA 94132}

\begin{abstract}
In Zombie Vortex Instability (ZVI), perturbations excite critical layers in stratified, rotating shear flow (as in protoplanetary disks), causing them to generate vortex layers, which roll-up into anticyclonic zombie vortices and cyclonic vortex sheets.  The process is self-sustaining as zombie vortices perturb new critical layers, spawning a next generation of zombie vortices. Here, we focus on two issues: the minimum threshold of perturbations that trigger self-sustaining vortex generation, and the properties of the late-time zombie turbulence on large and small scales.  The critical parameter that determines whether ZVI is triggered is the magnitude of the vorticity on the small scales (and not velocity); the minimum Rossby number needed for instability is $Ro_{crit}\sim0.2$ for $\beta\equiv N/\Omega = 2$, where $N$ is the Brunt-V\"{a}is\"{a}l\"{a} frequency. While the threshold is set by vorticity, it is useful to infer a criterion on the Mach number; for Kolmogorov noise, the critical Mach number scales with Reynolds number: $Ma_{crit}\sim Ro_{crit}Re^{-1/2}$.  In protoplanetary disks, this is $Ma_{crit}\sim10^{-6}$. On large scales, zombie turbulence is characterized by anticyclones and cyclonic sheets with typical Rossby number $\sim$0.3.  The spacing of the cyclonic sheets and anticyclones appears to have a ``memory'' of the spacing of the critical layers.  On the small scales, zombie turbulence has no memory of the initial conditions and has a Kolmogorov-like energy spectrum.  While our earlier work was in the limit of uniform stratification, we have demonstrated that ZVI works for non-uniform Brunt-V\"{a}is\"{a}l\"{a} frequency profiles that may be found in protoplanetary disks.
\end{abstract}


\section{INTRODUCTION} \label{sec:introduction}

\subsection{Background}

Without a doubt, we know that: (1) gas accretes inward through protoplanetary disks (PPDs) while angular momentum is transported outward through some combination of hydrodynamic and/or magnetohydrodynamic waves, instabilities, and turbulence; and (2) sub-millimeter dust particles coalesce to form super-kilometer size planetesimals, through some combination of collisional agglomeration and/or gravitational clumping.  The ubiquity and diversity of planetary systems imply that these processes are indeed robust; and yet, there persists uncertainty as to the exact nature of the relevant dynamical mechanisms. Theoretical research has progressed on two parallel tracks: purely hydrodynamic processes versus magnetohydrodynamic (MHD) processes.  See \citet{armitage2011review} and \citet{turner2014review} for comprehensive reviews.

\citet{balbus91} applied the magnetorotational instability (MRI) of \citet{velikhov59} and \citet{chandra60}, and demonstrated that magnetic fields can destabilize Keplerian shear, leading to turbulence and outward transport of angular momentum. However, there exist relatively dense, cool and nearly neutral ``dead zones'' in PPDs ($\sim$1-10 AU) that likely lack sufficient coupling between matter and magnetic fields \citep{turnerdrake2009,blaes94}, except perhaps in thin surface layers that have been ionized by cosmic rays or protostellar X-rays \citep{gammie96}.  A review of the substantial MRI in PPD literature is beyond the scope of this work; we refer readers to the relatively recent review \citet{turner2014review} on transport and accretion processes in PPDs.   We do note that recent work has moved far beyond ideal MHD to include non-ideal effects such as the Hall term and ambipolar diffusion, both which seem to make MRI-driven turbulence less effective in dead zones \citep{kunzlesur2013,baistone2011}.

Convective overstability (ConO) and the vertical shear instability (VSI) have gained attention on the purely hydrodynamic front.  In ConO, radial entropy gradients that would be stable according to the Solberg-H\o iland criterion in the adiabatic limit may yet be unstable in the limit of efficient thermal relaxation \citep{klahrhubbard2014,lyra2014}.  The chief obstacles for ConO are that the cooling time must be relatively short, $\Omega\tau_{cool}\sim 1$,  and the radial entropy gradient must be negative so that $0<-N_r^2<\Omega^2$, where $N_r$ is the radial Brunt-V\"{a}is\"{a}l\"{a} frequency.   The latter constraint requires a disk surface density profile that is significantly flatter than most standard models.  In VSI, vertical shear induced by radial gradients of temperature (\eg a thermal wind, a baroclinic effect) that would otherwise be stable to the Kelvin-Helmholtz Instability (KHI) in the adiabatic limit may yet be unstable in the limit of rapid thermal relaxation \citep{umurhan2016,barker2015,stoll2014,nelson2013,urpin2003}.  However, the cooling times must be especially short: $\Omega\tau_{cool}\sim h$, where $h=H/r$ is the aspect ratio of the disk \citep{lin2015}.

\subsection{Previous Work on the Zombie Vortex Instability}

In \citet{barranco05}, while we were trying to develop models for 3D vortices in the midplanes of PPDs, we serendipitously discovered that the stratified regions above and below the midplane rapidly filled with anticyclonic vortices and cyclonic vortex sheets.  At the time, we hypothesized that internal gravity waves propagated away from the midplane and deposited their energy in stratified regions where the shear rolled vorticity perturbations into new vortices.  However, our original explanation was not entirely complete.  \citet{MPJH13}, hereafter \citetalias{MPJH13}, correctly diagnosed the true mechanism for the formation of these vortices and in the process identified a new purely hydrodynamic instability, which we now call the ``Zombie Vortex Instability" or ZVI.   In order to get at the essential nature of the phenomenon, \citetalias{MPJH13} stripped out complicating features of a protoplanetary disk (\eg spatially varying gravity and Brunt-V\"{a}is\"{a}l\"{a} frequency) and numerically investigated simple Couette flow with constant gravity and constant Brunt-V\"{a}is\"{a}l\"{a} frequency in the limit of the Boussinesq approximation.  In this far simpler system, \citetalias{MPJH13} found that a small perturbing vortex could trigger an instability in a rapidly-rotating, strongly stratified flow, yielding vigorous, space-filling vortices, vortex layers and turbulence.

\begin{figure}[ht]
\epsscale{1.0}
\plotone{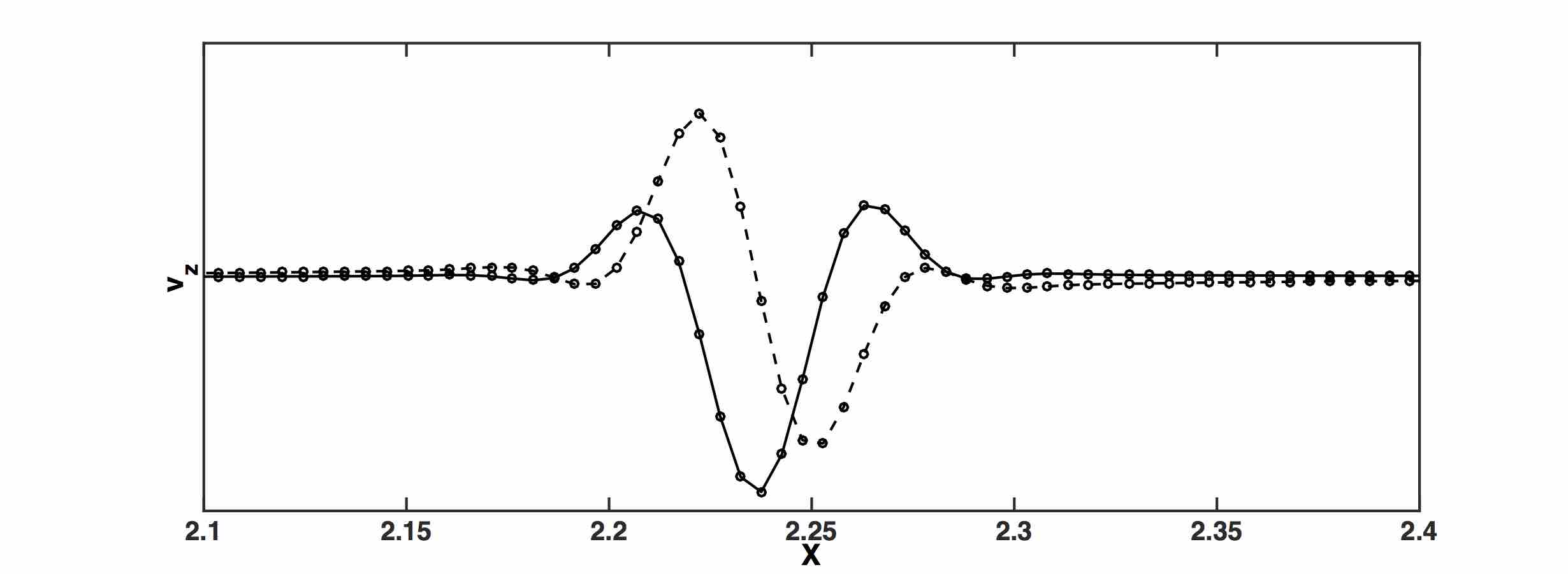}
\caption{\label{fig:velocityeigenmode}
Vertical velocity component (real \& imaginary parts) of the neutrally-stable critical-layer eigenmode $\boldsymbol{\hat{v}}(x)\exp[i(k_y y + k_z z - s t)]$ that leads to ZVI.  Eigenmode was computed with a linear eigensystem solver. Vertical velocity coupled with rapid rotation around the vertical axis causes vortex stretching and the intensification of vorticity, creating vortex layers at the location of the critical layers.  The unit of length is $H_0$, and the unit of velocity is arbitrary.  The purpose of the figure is to show that the eigenmode has near-zero amplitude everywhere except within the thin critical layers.  Viscosity was included to show that the critical layer can be numerically resolved; the Reynolds number was $Re\sim10^7$. The eigenmode here was computed with the Boussinesq equations with $N_0/\Omega_0 =2$, $k_y L_y = k_z L_z = 2 \pi$, and with the boundary condition $\hat{v}_x =0$ at $|x| = 4$, rather than shearing box boundary conditions, which do not have temporal eigenmodes. The number of Chebyshev modes to compute the eigenmode in the $x$ direction was $N_x=2048$.}
\end{figure}

The crucial new insight in \citetalias{MPJH13} was the recognition of ``baroclinic critical layers'' as being sites that are receptive to perturbations.  Critical layers are special locations in a shear flow where the coefficients of the highest derivatives of the linearized equations vanish, indicating that the neutrally-stable eigenmodes are singular there \citep{maslowe86, drazinreid81}.  Despite the singularities, critical layers are not spurious numerical solutions, but are true physical feature within the flow.  In the presence of dissipation in the form of (hyper)viscosity or thermal (hyper)diffusivity, or with the inclusion of nonlinear effects,  the eigenmodes are no longer truly singular, but still retain highly-localized regions where the gradients of the density, pressure and velocity can be extremely large  (Fig.~\ref{fig:velocityeigenmode}).

In \citet{MPJBHL2015}, hereafter \citetalias{MPJBHL2015}, we presented a cartoon model for ZVI (see Fig.~1 in that paper).  Initial perturbations (either a vortex or noise with a power-law energy spectrum) excite baroclinic critical layers.  These critical layers then generate dipolar vortex layers (two juxtaposed oppositely-signed layers of vorticity); while cyclonic vortex layers remain stable, anticyclonic vortex layers roll up into anticyclonic vortices (\ie anticyclones).  The crux of ZVI is that these new zombie vortices then ``infect'' neighboring critical layers with perturbations, which generate new vortex sheets, which spawn new zombie vortices -- and this self-sustaining process continues unabated until the dead zone is filled with zombie vortices.  The instability is not an artifact of the numerical method as we have observed it with spectral codes and finite-volume codes (\eg {\it Athena}), with fully compressible, anelastic and Boussinesq treatments of the continuity equation, with and without the shearing box, and with either hyperviscosity or real molecular viscosity.  We believe ZVI was not observed in many earlier numerical studies because they were missing one of the necessary ingredients: vertical stratification, high resolution to resolve the narrow critical layers, a broad spectrum of perturbations (\ie Kolmogorov, but not Gaussian-peaked), and enough simulation time to allow the critical layers to amplify perturbations.

\subsection{Goals \& Outline}

ZVI is a subcritical (finite-amplitude) instability.  One of the chief goals of this work is to quantify the minimum threshold of perturbations to excite critical layers and trigger the onset of ZVI.  In \citetalias{MPJH13}, we instigated the onset of ZVI with a single perturbing vortex, whereas in \citetalias{MPJBHL2015} we investigated triggering ZVI with random noise.  Both work to kick-off the instability, but using a single vortex has a couple advantages: first, it is far easier to observe the excitation of individual critical layers; second, one can more readily characterize the strength of the perturbation and initialize a vortex with a well-defined vorticity.  We have found, numerically, that a perturbing vortex can excite neighboring critical layers, creating localized regions of vertical velocity.   In a fluid rotating rapidly around the $z$-axis, vertical velocity causes ``vortex stretching'' and the intensification of vorticity \citep{pedlosky79}.  The newly-created vorticity is then stretched by the shear, forming dipolar vortex layers.  We observe that a weaker perturbing vortex (as measured by the strength of vorticity) yields weaker vortex layers, and a stronger perturbing vortex yields stronger vortex layers.  It is well known that vortex layers with sufficient strength of their vorticity are linearly unstable to the development of waves which break and roll-up into discrete vortices \citep{marcus90a,marcus93}.  In our numerical experiments, we find that the strength of the initial perturbing vortex must exceed a certain threshold to yield vortex layers that are strong enough to become linearly unstable.  When this occurs, the resultant first generation of zombie vortices become a source for new perturbations that excite neighboring critical layers.  However, this does not always yield a run-away process.  We have observed cases in which the first generation of zombie vortices may be weaker than the original perturbing vortex, enough so that strength of the perturbing vorticity is below the threshold to create sufficiently strong vortex layers can go linearly unstable, halting the further development of ZVI.  On the other hand, if the strength of the vorticity in the original perturbing vortex is above some higher threshold, we observe in numerical experiments that the first generation of zombie vortices can be just as strong, exciting new critical layers and yielding vortex layers that are strong enough to be linearly unstable to roll-up into a next generation of zombie vortices, and so on.  We say that the flow {\it zombifies} when this process becomes self-sustaining, filling the domain with zombie vortices and non-isotropic zombie turbulence.

Our numerical experiments revealed that the threshold of the initial perturbation needed to trigger ZVI depend on the magnitude of the vorticity of the initial vortex, rather than on its velocity (which could be varied while maintaining a fixed value of vorticity by changing the diameter of the initial vortex), or any other feature that we could identify.  However, it is still unclear whether this remains true when the initial perturbations are some spectrum of random noise (\eg turbulence with a Kolmogorov energy spectrum) rather than a coherent vortex.  Is it still the magnitude of the vorticity (or a dimensionless form of the vorticity such as the Rossby number $Ro$) that determines whether or not ZVI is triggered, or does the initial velocity, energy, or some other property of the initial noise determine the onset of instability?  What are the threshold values of $Ro$, or of the velocity (or the  dimensionless expression of the velocity, such as its Mach number $M\!a$ or Reynolds number $Re$) of the noise when ZVI is triggered?

The second goal of this paper is to characterize the properties of fully-developed zombie turbulence. By doing so we can distinguish zombie turbulence from other forms of turbulence and illustrate its unique properties -- in particular, those that might contribute to its ability to transport angular momentum, to concentrate or mix dust, and to disrupt other dynamics that might occur in a fully laminar PPD.  We shall show that zombie turbulence is far from laminar or weakly chaotic and has many properties in common with fully-developed, homogeneous turbulence. However, zombie turbulence is not isotropic, and unlike other forms of turbulence it has a ``memory'' of how it formed. The ``memory'' is not of its initial conditions, but rather of the linear eigenmodes responsible for triggering the instability. Those eigenmodes, especially, their structure in the radial direction remain imprinted on the flow indefinitely and lead to persistent anticyclonic vortices and cyclonic layers at large length scales and a more classic Kolmogorov turbulence at all other lengths. 

The outline of the remainder of this paper is as follows. In \S~2, we present the hydrodynamic equations and offer a brief review of basic turbulence concepts.  In \S~3, we describe the results of a series of numerical experiments focused on elucidating what triggers ZVI, and demonstrate that it is the value of the Rossby number of the initial noise, rather than its Mach number or energy that determines whether ZVI is triggered. In \S~4 we investigate the properties of space-filling zombie turbulence on large and small scales and show how it differs from other forms of turbulence.  A summary and future work appear in \S~5.

\section{Hydrodynamic Equations \& Brief Review of Turbulence Concepts}

\subsection{Equations of Motion \& Steady-State Background}\label{subsec:equations}

Consider a three-dimensional box located at cylindrical radius $R_0$ from the protostar that co-rotates with the gas with the Keplerian angular rate $\Omega_0\equiv\Omega_K(R_0)$.  The box is sufficiently small that we ignore curvature and choose Cartesian coordinates $(x,y,z)$ for the local radial, azimuthal, and vertical directions, respectively \citep{hill1878, goldreich65b}.  Corresponding unit vectors are $\boldsymbol{\hat{x}}$, $\boldsymbol{\hat{y}}$, and $\boldsymbol{\hat{z}}$.  Stratification is measured by the Brunt-V\"{a}is\"{a}l\"{a} frequency $N(z) \equiv \sqrt{(g/C_P)(d\bar{s}/dz)}$, where $g$ is the vertical component of the acceleration of gravity, $C_P$ is the specific heat at constant pressure, and $\bar{s}(z)$ is the vertical entropy profile.  As in \citetalias{MPJBHL2015},  we limit our study to flows with spatially-uniform vertical stratification, so we choose a uniform background temperature $T=T_0$ and constant acceleration of gravity $g=g_0$,  which yields a constant Brunt-V\"{a}is\"{a}l\"{a} frequency $N_0 = g_0/\sqrt{C_P T_0}$.  The steady-state equilibrium of the stratified, rotating, sheared flow is thus:
\begin{subequations}\label{E:equilibrium}
\begin{align}
\boldsymbol{\bar{v}}(x)  &= \left(\bar{v}_x,  \bar{v}_y, \bar{v}_z\right) = \left(0, -3\Omega_0 x/2, 0\right), \label{E:shear} \\
\bar{\rho}(z) &= \bar{P}(z)/\mathcal{R}T_0 =  \rho_0\exp(-z/H_0), \label{E:hydrostatic}
\end{align}
\end{subequations}
where $\boldsymbol{v}$ is the velocity in the rotating frame, $\rho$ is gas density, $P$ is gas pressure, $\rho_0$ is the equilibrium density at the disk midplane, $H_0\equiv\mathcal{R}T_0/g_0$ is the vertical pressure scale height, and $\mathcal{R}$ is the gas constant.  Overbars are used to indicate equilibrium steady-state variables.  We model the temporal evolution of the flow with the Euler equations with the continuity equation replaced by the anelastic approximation and a linearized ideal gas law:
\begin{subequations}\label{E:anelastic}
\begin{align}
0 &= \boldsymbol{\nabla} \cdot [\bar{\rho}(z)\boldsymbol{v}], \label{E:anelastic1}\\
{\frac{\partial\boldsymbol{v}}{\partial t}}  &= -(\boldsymbol{v}\cdot \boldsymbol{\nabla})\boldsymbol{v}-2\Omega_0\boldsymbol{\hat{z}}\times\boldsymbol{v} +3 \Omega_0^2 \, x\boldsymbol{\hat{x}}\nonumber\\
&{}-\boldsymbol{\nabla}\left[\frac{{P - \bar{P}(z)}}{{\bar{\rho}(z)}}\right] + \left[{\frac{T - T_0}{T_0}}\right] g_0\boldsymbol{\hat{z}}, \label{E:anelastic2}\\
{\frac{\partial T}{\partial t}} &= - (\boldsymbol{v} \cdot \boldsymbol{\nabla}) T  - \left(\frac{N_0^2}{g_0}\right)Tv_z, \label{E:anelastic3}\\
\left[{\frac{P - \bar{P}(z)}{\bar{P}(z)}}\right] &=\left[{\frac{\rho - \bar{\rho}(z)}{\bar{\rho}(z)}}\right] + \left[{\frac{T - T_0}{T_0}}\right].\label{E:anelastic4}
\end{align}
\end{subequations}
The anelastic approximation has been extensively used in the study of deep, subsonic convection in planetary atmospheres \citep{ogura62,gough69,bannon96} and stars \citep{gilman81,glatzmaier81a,glatzmaier81b}. We have previously used the anelastic approximation to study three-dimensional vortices in PPDs \citep{barranco00a,barranco05,barranco06} and the Kelvin-Helmholtz instability of settled dust layers in PPDs \citep{barranco09,lee2010a,lee2010b}.  The basic idea is that there may be large variations in the background pressure and density in hydrostatic equilibrium, but that at any height in the atmosphere, the fluctuations of the pressure and density are small compared to the background values at that height.

We use shearing box boundary conditions \citep{goldreich65b,marcus77,rogallo81}.  In the vertical direction, we use periodic boundary conditions (rather than rigid lid boundaries) because it is easier to analyze energy spectra in Fourier space.  In our computations, we choose $M\ge256$ Fourier modes in each of the three spatial directions.  Generally, $M$ indicates the amount of spatial resolution, and if a numerical experiment is sensitive to the value of $M$, we are generally skeptical of the results and assume the simulation is under-resolved. However, $M$ has a physical meaning, not just a numerical one.  As shown below, $M$ is a measure of the ``length'' of the turbulent spectrum in wavenumber space, and this length represents a physical quantity that turns out to be relevant to whether or not ZVI is triggered. In a simulation of Kolmogorov turbulence, the ratio of the largest length (the production range of the self-similar inertial part of the energy spectrum) to the smallest length (the dissipation or Kolmogorov length), is  $Re^{3/4}$, where $Re$ is the Reynolds number \citep{tennekes1972first}. In the  ``dissipationless'' numerical simulations presented here, the flows are computed with a hyperviscosity and hyperdiffusivity to stabilize the calculations and the effective ratio of the largest to smallest length scale in the numerical computations, as well in the initial turbulence, is $M/2$. 

\subsection{Review of Turbulent Spectra, Eddy Velocities, Eddy Vorticities, and Fourier Modes} \label{subsec:eddies}

To better understand how ZVI is triggered from initial noise, we now review the nomenclature and ideas used in describing homogeneous, isotropic turbulence with no spatial structures or correlations (which is how we define initial ``noise'' in this paper).  To simplify our analysis and avoid unnecessary confusion, we restrict this discussion to incompressible turbulence (which is not an unreasonable approximation when there is not a large variation in the value of $\bar{\rho}(z)$ in the computational domain).  The differential kinetic energy spectrum $E(k)$ as a function of spatial wave number $k\equiv|\boldsymbol{k}| = \sqrt{k^2_x+k^2_y+k^2_z}$ is 
\begin{equation}
dE = (1/2)|\boldsymbol{v}(\boldsymbol{r})|d^3\boldsymbol{r} = E(k)dk,
\end{equation}
where the velocity in a periodic cubic box of size $L^3$ is written as a discrete sum of Fourier modes: 
\begin{equation}
\boldsymbol{v}(\boldsymbol{r}) = \sum_{n_x} \, \sum_{n_y} \, \sum_{n_z} \, \boldsymbol{\tilde{v}}_{\boldsymbol{k}}\, e^{i \, \boldsymbol{k} \cdot\boldsymbol{r}}, \label{turbulence1}
\end{equation}
where $k_j \equiv n_j\Delta k$ for $j=$ $x$, $y$, $z$, $\Delta k\equiv 2\pi/L$, and $n_j$ are integers in the interval $[-M/2, +M/2]$.  The condition that $\boldsymbol{v}(\boldsymbol{r})$ is real implies $\boldsymbol{\tilde{v}}_{-\boldsymbol{k}}=\boldsymbol{\tilde{v}}_{\boldsymbol{k}}^*$, where $()^*$ indicates complex conjugation.  Often, the spectrum has a power-law dependence on $k$, so $E(k)=E_0k^{-a}$ with normalization constant $E_0$ and spectral index $a$.  For example, Kolmogorov turbulence has $a=5/3$.  Moments of the energy spectrum yield useful quantities: the total kinetic energy is trivially the zeroth moment $E \equiv \int_0^{\infty}E(k)dk$, while enstrophy is the second moment $E_2\equiv\int_0^{\infty}E(k)k^2dk$.  From these moments, we can define rms velocity $v_{rms} \equiv \sqrt{2E}$, rms Mach number $M\!a_{rms}\equiv v_{rms}/C_s$, rms vorticity $\omega_{rms}\equiv\sqrt{2E_2}$, and rms Rossby number $Ro_{rms}\equiv\omega_{rms}/(2\Omega_0)$.  If we consider only the vertical component of vorticity, then $Ro_{z,rms}=Ro_{rms}/\sqrt{3}$ for homogeneous, isotropic turbulence.

It is instructive to think of turbulence as a sequence of eddies in which the diameter of an eddy in the sequence is equal to half the diameter of the preceding eddy in the sequence \citep{tennekes1972first}.   An eddy with wavenumber $k$ and
length scale $\ell\equiv 2 \pi/k$ has  kinetic energy  $\int_{k/2}^{k} E(k') dk'$ and has an rms eddy velocity of
$V_{eddy}(\ell) = [2 \int_{k/2}^{k} E(k') dk']^{1/2}$. We relate the eddy Rossby number $\widetilde{Ro}(k)$ and the eddy Mach number $\widetilde{M\!a}(k)$ by:
\begin{equation}
\widetilde{Ro}(k) \equiv V_{eddy}(\ell)/(2 \Omega_0\ell)  \equiv C_s\widetilde{M\!a}(k) /(2 \Omega_0\ell).\label{E:RoMa}
\end{equation}
For an energy spectrum $E(k)$ with spectral index $a$,
\begin{subequations}
\begin{align}
V_{eddy}(\ell) &= V_{eddy}(L) \left[\frac{\ell}{L}\right]^{(a-1)/2}, \label{E:eddy_scale1} \\
\widetilde{M\!a}(k) &\propto k^{(1-a)/2},\label{E:Ma_scale1}\\
\widetilde{Ro}(k) &\propto k^{(3-a)/2}.\label{E:Ro_scale1}
\end{align}
\end{subequations}
For a Kolmogorov spectrum with $a=5/3$, 
\begin{subequations}
\begin{align} 
V_{eddy}(\ell) &= V_{eddy}(L) \left[\frac{\ell}{L}\right]^{1/3}, \label{E:eddy_scale2} \\
\widetilde{M\!a}(k) &\propto k^{-1/3}, \label{E:Ma_scale2} \\
\widetilde{Ro}(k) &\propto k^{2/3}. \label{E:Ro_scale2}
\end{align}
\end{subequations}
For turbulence with spectral index $1 < a < 3$, eddy velocity and kinetic energy decrease with decreasing length scale, while eddy vorticity and enstrophy increase. Equivalently, with increasing $k$, $\widetilde{Ro}(k)$ increases and $\widetilde{M\!a}(k)$ decreases. The implication of this is that most of the kinetic energy is at the large length scales and that the largest eddies contribute the most to the rms Mach number, while most of the enstrophy is at the smallest length scales and that the smallest eddies contribute most to the rms Rossby number.  For a turbulent spectrum with a large inertial range (\ie the ratio of largest to smallest wavenumbers is big), the ratio of the rms velocity of the largest eddies to the rms velocity of the total flow is 
\begin{equation}
V_{eddy}(L)/v_{rms} = [1- (1/2)^{(a-1)}]^{1/2}, \label{E:Ma_scale3}
\end{equation}
so for Kolmogorov turbulence, the rms Mach number of the largest eddy is $\sim60\%$ that of the rms Mach number of the total flow.  If the smallest length scale of the turbulence is set by viscous dissipation, then that length is $\ell_{\nu} \equiv  \nu/V_{eddy}(\ell_{\nu})$, where $\nu$ is the kinematic viscosity, and Eq.~(\ref{E:eddy_scale1}) shows that 
\begin{equation}
\ell_{\nu}/L = Re^{-2/(a+1)}, \label{E:reynolds}
\end{equation}
where $Re \equiv [LV_{eddy}(L)]/\nu$ is the Reynolds number of the flow.  For Kolmogorov turbulence, $\ell_{\nu}$ is called the Kolmogorov length and is equal to $Re^{-3/4}L$.

\begin{figure}[ht]
\epsscale{0.45}
\plotone{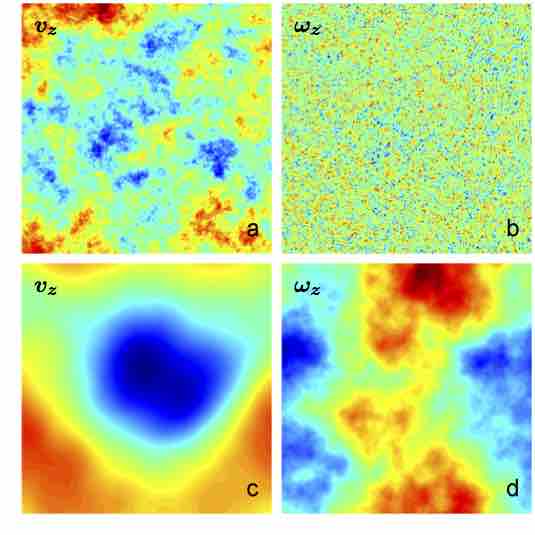}
\caption{\label{fig:turbspec}
Vertical velocity (panels a and c) and vertical vorticity (panels b and d) in the $x$-$y$ plane for Kolmogorov noise with spectral indices $a=5/3$ (panels a and b) and $a=5$ (panels c and d).  The color-map ranges from blue (positive) to red (negative) with green as zero.  For $a=5/3$, the largest length scales dominate the velocity, while the smallest length scales dominate the vorticity.  For $a=5$, both the velocity and vorticity are dominated by the largest length scales.}
\end{figure}

It is crucial to note that an eddy is not equivalent to a Fourier mode $\boldsymbol{\tilde{v}}_{\boldsymbol{k}}$ of the velocity field, as defined in Eq.~(\ref{turbulence1})\footnote{Unfortunately, there has been some confusion in the astrophysics literature that incorrectly states that $|\boldsymbol{\tilde{v}}_{\boldsymbol{k}}|$ scales with $k$ in the same way that $V_{eddy}$ does, which is not true.}. Rather, an eddy is defined by the sum or integral of a band of Fourier modes with different wavenumbers near wavenumber $|\boldsymbol{k}|$.  For example, consider eddies containing wavenumbers between $k$ and $2k$; there are $(L/2\pi)^3\int_k^{2k}4\pi k'^2dk' = 7(4 \pi/3)(Lk/2\pi)^3$ Fourier modes in this band.  Equate the energy in these Fourier modes with the eddy kinetic energy: $[7(4 \pi/3)(Lk/2\pi)^3]|\boldsymbol{\tilde{v}}_{\boldsymbol{k}}|^2 = (1/2)[V_{eddy}(2 \pi/k)]^2$, which yields:
\begin{equation}
|\boldsymbol{\tilde{v}}_{\boldsymbol{k}}| \propto k^{-(a+2)/2}.   \label{turbulence2}
\end{equation}
The velocity of the initial noise used in the calculations in this paper were created using Eqs.~(\ref{turbulence1}) and~(\ref{turbulence2}) where the $\boldsymbol{\tilde{v}}_{\boldsymbol{k}}$ have random phases.  Fig.~\ref{fig:turbspec} illustrates how the spectral index $a$ affects the spatial pattern and length scales of the vertical velocity and vertical vorticity of the initial noise in our calculations.  The first row of figures show noise with spectral index $a=5/3$, demonstrating that the largest length scales dominate the velocity, while the smallest length scales dominate the vorticity.  In the second row, the spectral index is $a=5$, and both the velocity and vorticity are dominated by the largest length scales.

\section{Numerical Experiments to Elucidate the Trigger for ZVI}\label{sec:thresholds}

\begin{table}[h]
\centering
\caption{Summary of Numerical Experiments in \S\ref{sec:thresholds}} \label{tab:experiments}
\begin{tabular}{llll}
\hline
\hline
Set & Resolution $M$ & Spectral Index $a$ & Spectral Magnitude $E_0$\\
\hline
1 & 256	& 5/3		& vary\\
2 & 256	& vary	& vary (holding $\int E(k)dk$ fixed)\\
3 & vary	& 5/3		& fixed\\
4 & vary	& 5/3		& vary\\
\hline
\end{tabular}
\end{table}

One of our goals is to disentangle the exact nature of the trigger for ZVI. What matters most: the peak perturbation velocity, or the non-Keplerian kinetic energy, or the peak perturbation vorticity, or non-Keplerian enstrophy?  We shall separate the non-Keplerian part of the flow from the Keplerian differential rotation and define perturbation velocity $\boldsymbol{\tilde{v}}$, non-Keplerian kinetic energy $KE$, relative vorticity $\boldsymbol{\tilde{\omega}}$, and point-wise Rossby number of the vertical component of the relative vorticity $Ro(x,y,z,t)$:
\begin{subequations}
\begin{align}
\boldsymbol{\tilde{v}} &\equiv \boldsymbol{v} - \boldsymbol{\bar{v}},\\
KE &\equiv (1/2)\int \bar{\rho}|\boldsymbol{\tilde{v}}|^2~dV,\\
\boldsymbol{\tilde{\omega}} &\equiv \boldsymbol{\nabla} \times \boldsymbol{\tilde{v}},\\
Ro(x, y, z, t) &\equiv \tilde{\omega}_z(x, y, z, t)/(2\Omega_0).\label{E:rossby_number}
\end{align}
\end{subequations}
We have found that initial noise with a wide range of parameter values can trigger ZVI.  Eqs.~(\ref{E:anelastic}) can be rewritten in dimensionless form such that they contain only five dimensionless parameters: $\beta\equiv N_0/\Omega_0$, $\gamma\equiv C_P/C_V$, $L_x/H_0$, $L_y/H_0$, and $L_z/H_0$, where the computational box is of size $L_x \times L_y \times L_z$.  As in \citetalias{MPJBHL2015}, we set $\beta=2$, $\gamma=5/3$ and $L_x/H_0 = L_y/H_0 = L_z/H_0 =1$.   In the following sets of numerical experiments, we will keep the stratification fixed, and only vary properties of the energy spectrum $E(k)=E_0k^{-a}$ of the initial noise, or the resolution $M$.  

All the simulations in this work use a suite of related codes all based on the original pseudo-spectral code of \citet{barranco06}, which has specially tailored algorithms to handle shear, rotation and stratification.  This code has been used in our previous related studies of vortex dynamics in PPD and the stability of dust layers to Kelvin-Helmholtz Instability (KHI) \citep{barranco05,barranco09,lee2010a,lee2010b,MPJH13,MPJBHL2015}.

\subsection{First Set of Experiments --  Effect of Overall Magnitude of Energy Spectrum, $E_0$}

\begin{figure}[ht]
\epsscale{1.0}
\plotone{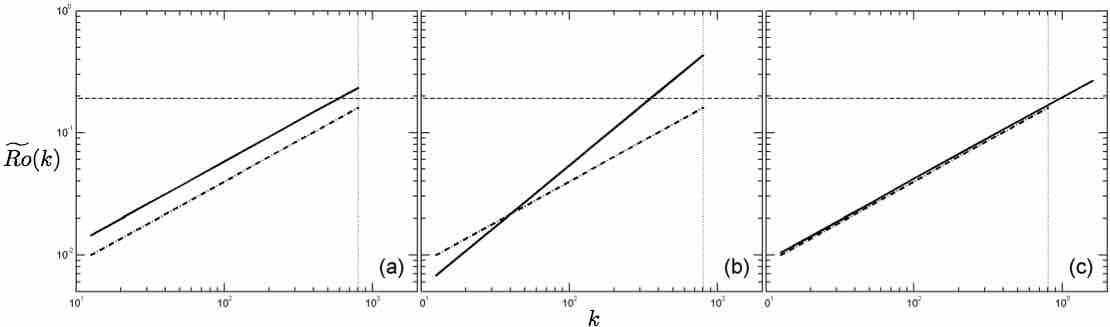}
\caption{\label{fig:rossby}
Vorticity spectra $\widetilde{Ro}(k)$ of the initial noise as a function of wavenumber $k$ (in units of $L^{-1}$), for the set of numerical experiments in \S\ref{sec:thresholds} to elucidate the trigger for ZVI.  $\beta\equiv N_0/\Omega_0=2$ for numerical experiments in this section.  {\bf In all panels:} The thin horizontal dashed line is $\widetilde{Ro}=0.19$, which corresponds to the threshold Rossby number $\widetilde{Ro}(k_{res})$ of the initial Kolmogorov noise that triggers zombie turbulence for $M=256$. The thin vertical dotted line in all panels indicates the resolution wavenumber $k_{res} =256\pi\approx 800$, which is the resolution used in the first two sets of experiments. In all panels, the thick dashed line with slope $2/3$ represents the vorticity spectrum $\widetilde{Ro}(k)$ of the reference run, which did not exhibit ZVI. ({\bf a}) First set of experiments. The energy spectra of the initial noise was Kolmogorov with $a=5/3$; only $E_0$ was varied.  Not all initial conditions are shown, but instead we show one representative initial condition: the thick solid line corresponds to an initial condition with $\widetilde{Ro}(k_{res})>0.19$; this initial condition triggered ZVI.  ({\bf b}) Second set of experiments. The spectral index $a$ was varied, while the value of $E_0$ was chosen to keep the total kinetic energy of the initial noise fixed at the same value as in the reference run.  Not all initial conditions are shown, but instead we show one representative initial condition: the thick solid line corresponds to spectral index $a=1$; this case triggered ZVI, consistent with our hypotheses that the criterion for instability is $\widetilde{Ro}(k_{res})>0.19$. ({\bf c}) Third set of experiments. The resolution $k_{res}$ was varied, holding $a$ and $E_0$ constant.  The thick solid straight line corresponds to $M=512$ and $k_{res} \approx 1600$; this case triggered ZVI, consistent with our hypotheses that the criterion for instability is $\widetilde{Ro}(k_{res})>0.19$.}
\end{figure}

Fig.~\ref{fig:rossby} summarizes the initial conditions for a series of numerical experiments to determine what triggers ZVI.  First, we computed a reference run with a Kolmogorov spectrum ($a=5/3$) of noise with low initial amplitude $E_0$ specifically chosen to be somewhat {\it below} the threshold to trigger ZVI; the initial non-Keplerian kinetic energy of this stable reference run simply decayed in time.    The vorticity spectrum $\widetilde{Ro}(k)$ for this stable reference run is illustrated by the thick dashed line in all three panels of Fig.~\ref{fig:rossby}.  Consistent with Eq.~(\ref{E:Ro_scale2}), $\widetilde{Ro}(k)$ increases with wavenumber as $k^{2/3}$ (and thus appears as a line with slope of 2/3 in a log-log plot), while $\widetilde{M\!a}(k)$ (not plotted) decreases as $k^{-1/3}$. The reference run had $M=256$ Fourier modes in each spatial dimension and the resolution wavenumber (\ie maximum wavenumber included in simulation) was $k_{res} \equiv \pi M/L \approx 800$.  The vertical dotted line in all three panels of Fig.~\ref{fig:rossby} shows the resolution wavenumber of this reference run.

Keeping spectral index $a=5/3$ fixed and spectral resolution $M=256$ fixed, we varied only the overall magnitude of the Kolmogorov energy spectrum $E_0$.  The goal here was not to pin down the exact triggering amplitude to any sort of high precision, but more so to establish a baseline run to see which characteristics of the initial perturbations are critical to ZVI.  The initial vorticity spectra $\widetilde{Ro}(k)$ for this set of runs appear as parallel lines to the thick dashed line of the reference run in Fig.~\ref{fig:rossby}; lines above have a greater $E_0$, and lines below have a smaller $E_0$.

As expected, all initial conditions with $E_0$ smaller than that of the reference run failed to destabilize the flow, but a sufficiently larger value of $E_0$ triggered ZVI.   In the left panel in Fig.~\ref{fig:rossby}, one can see that Increasing $E_0$, holding spectral index $a$ fixed, causes the entire vorticity spectrum to shift upward, resulting in the peak vorticity at the smallest scales to exceed some critical value.  While we found a critical value of $E_0$ for Kolmogorov noise that results in triggering ZVI at this resolution, our working hypothesis is that this is not the relevant parameter for determining the onset of instability; rather, we believe it is the peak vorticity or Rossby number that is the discriminant.  For the run with the Kolmogorov spectrum with the minimum amplitude to trigger ZVI, we determined the peak vorticity, which is on the smallest resolved length scale: $\widetilde{Ro}(k_{res})=0.193$.  This value is plotted as a thin horizontal dashed line in all three panels of Fig.~\ref{fig:rossby}. 

\subsection{Second Set of Experiments -- Effect of Spectral Index, $a$}

We now hypothesize that the requirement for triggering ZVI is that the initial noise have peak Rossby number greater than some threshold level.  For $1<a<3$, the peak Rossby number occurs at the smallest scales, so the criterion may be $\widetilde{Ro}(k_{res})\gtrsim0.2$.  In this next set of experiments, we want to increase the vorticity on the smallest scales without increasing the total energy or the vorticity and velocity on the largest scales.   We hold the resolution fixed $M=256$ ($k_{res} \approx 800$), and we vary spectral index $a$ while normalizing the energy spectrum so that the total perturbation kinetic energy $E = \int E(k) dk$ is kept at the same value as the reference run.  

In Fig.~\ref{fig:rossby}(b), the thick solid line corresponds to spectral index $a=1$.  The peak Rossby number at the smallest scales exceeds 0.2, and the noise triggers ZVI, producing sustained zombie turbulence.  Using binary chop, we determined the threshold value of $a$ needed to trigger ZVI.  However, this value of $a$ is {not} important or universal, but will depend on the spectral resolution $k_{res}$ and the overall magnitude of the noise $E_0$.  If one decreased the overall magnitude of noise, one could always decrease (within reason) the spectral index so that the peak vorticity at smallest scales exceeds $\sim$ $0.2$.

Thus, this second set of numerical experiments is consistent with, but does not yet definitively prove, our hypothesis that the required condition for noise to trigger ZVI is determined by $\widetilde{Ro}(k_{res})$.  In addition, this set of experiments proves that the necessary condition for triggering ZVI does not uniquely depend upon the amplitude of the kinetic energy of the initial noise (or, equivalently, its rms Mach number) because the initial energy of the noise was the same in all of these experiments.  This is an important finding because it appears that the necessary condition to trigger instability in many other finite-amplitude unstable flows is set by the energy of the initial perturbation.

\subsection{Third Set of  Experiments -- Effect of Resolution, $k_{res}$}

In the third set of numerical experiments, we fix the spectral index $a=5/3$ and the overall magnitude of the noise $E_0$ to have the same value as in the stable reference run, which had a peak vorticity at the resolution scale below 0.193.  What would happen if we increased the extent of the inertial range by increasing the spectral resolution?  Because the vorticity spectrum increases as $k^{2/3}$, including higher wavenumbers will yield higher peak vorticity on the smallest scales, without changing the velocity and vorticity on the largest scales.  The simulation corresponding to the solid line in Fig.~\ref{fig:rossby}(c) has $M=512$, $k_{res} \approx 1600$, and produces sustained zombie turbulence via ZVI.  By carrying out a binary chop search on $k_{res}$ between 800 and 1600, we found the threshold value of $k_{res}$ that produces zombie turbulence.   However, this critical value of  $k_{res}$ is not important or universal, but will depend on the overall magnitude of the noise $E_0$ and the spectral index $a$.  With larger values of $E_0$ or $a$, ZVI can be triggered with a smaller inertial range or lower resolution.  

Thus, this third set of numerical experiments is consistent with, but does not yet definitively prove, our hypothesis that the required condition for noise to trigger ZVI is determined by $\widetilde{Ro}(k_{res})$.  In addition, this set of experiment proves that the necessary condition for triggering ZVI does not uniquely depend upon the functional form of the kinetic energy spectrum $E(k) =E_0k^{-a}$ of the initial noise because all of the spectra in this set of experiments have the same values of $E_0$ and $a$.  Neither the Mach number nor the Rossby number of largest eddies in the initial noise determines whether ZVI is triggered because the values of these numbers were the same in this set of experiments.  Although the values of the kinetic energy of the initial noise was not explicitly fixed in the third set of experiments, it turns out that in practice the kinetic energy was nearly the same: the difference in the kinetic energies of runs with resolutions of $k_{res}$ and $k'_{res}$ is $\int_{k_{res}}^{k'_{res}}E_0k^{-5/3}dk$, which is negligible compared to the total energy of the initial noise, $\int_{2 \pi/L}^{k'_{res}}E_0k^{-5/3}dk$.

\subsection{Fourth Set of  Experiments -- Does the Threshold Peak Vorticity Depend on Resolution?}

One tantalizing implication of the third set of experiments is that at sufficiently high enough resolution, the peak vorticity will always exceed some threshold level and therefore trigger ZVI, no matter how weak the noise is initially, as measured by total energy, or Rossby number and Mach number of the large scale eddies.  The logic is as follows:  (1) $\widetilde{Ro}(k_{res})$ increases with $k_{res}$ for a spectrum with $1<a<3$, (2) the threshold value $[\widetilde{Ro}(k_{res})]_{crit}$ is invariant with respect to the value of $k_{res}$, and therefore (3) for large enough $k_{res}$, initial noise with a given value of $E_0$ will always go unstable to ZVI.  The caveat in this reasoning is that we have not yet shown that the threshold value $[\widetilde{Ro}(k_{res})]_{crit}$ is invariant with respect to the values of $k_{res}$. 

\begin{figure}[ht]
\epsscale{0.4}
\plotone{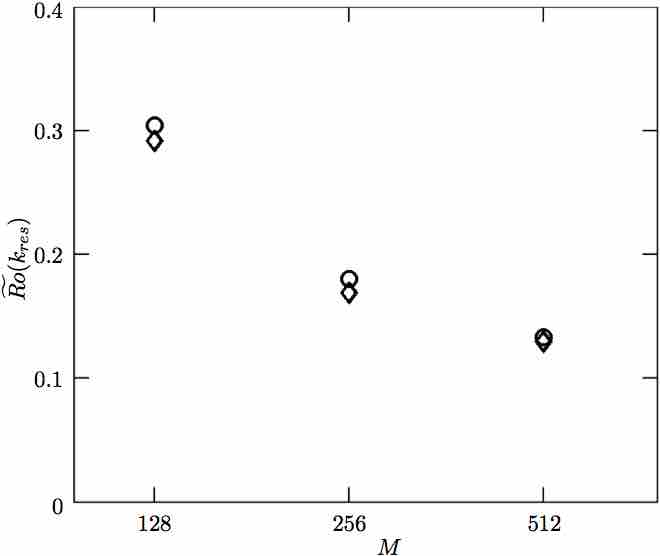}
\caption{\label{fig:exp4}
The threshold value of $\widetilde{Ro}(k_{res})$ as a function of $M \equiv L k_{res}/\pi$. $\beta\equiv N_0/\Omega_0=2$ and pectral index i$a=5/3$. For each value of $M$, a binary chop was carried out on the value of $E_0$ to determine the threshold value of $\widetilde{Ro}(k_{res})$. In other words, this is similar to the first set of numerical experiments, but carried out for varying resolution.  Circles show the values of $\widetilde{Ro}(k_{res})$ of the initial Kolmogorov noise that produced sustained zombie turbulence, while the diamonds shows the values of  $\widetilde{Ro}(k_{res})$ of the initial Kolmogorov noise that decayed. The threshold values of $\widetilde{Ro}(k_{res})$ are between the diamonds and circles for any given $M$.  It is not clear (nor important) whether $[\widetilde{Ro}(k_{res})]_{crit}$ asymptotically plateaus at some value at higher resolution; all that matters is that it is bounded from above.}
\end{figure}

The purpose of the fourth set of numerical experiments is to investigate the behavior of the threshold value $[\widetilde{Ro}(k_{res})]_{crit}$ with varying resolution. If the threshold value $[\widetilde{Ro}(k_{res})]_{crit}$ is invariant with respect to the value of $k_{res}$ {\it or} if it {\it decreases} with increasing $k_{res}$, then for large enough $k_{res}$ a flow with a given $E_0$ will always go unstable to ZVI.  In essence, we repeat the first set of experiments, but now at different resolutions. That is, we fix the spectral index $a=5/3$, then vary $E_0$ until we find the minimum value that triggers ZVI, and then record the eddy Rossby number at the resolution wavenumber.  We repeat at different resolutions to determine $[\widetilde{Ro}(k_{res})]_{crit}$ as a function of resolution.  Figure~\ref{fig:exp4} illustrates that, in fact, the threshold value $[\widetilde{Ro}(k_{res})]_{crit}$ decreases with increasing $k_{res}$ and that the critical value is less than or equal to 0.14 at the highest resolution we investigated.\footnote{It is possible that our numerical calculations with low values of $k_{res}$ may contain numerical errors due to our use of hyperviscosity.  Hyperviscosity artificially dissipates energy for wavenumbers near the resolution limit; flows with larger values of $k_{res}$ have a smaller fraction of their eddies dissipated by hyperviscosity and therefore are more accurate.}

We did not investigate higher resolutions because the computational costs were prohibitive, so we were not able to determine whether $[\widetilde{Ro}(k_{res})]_{crit}$ asymptotically approaches a plateau.  However, the precise value of the 
threshold value of $\widetilde{Ro}(k_{res})$ is not important, only that it is bounded from above. 

\subsection{Implications for Triggering ZVI in Protoplanetary Disks}

Our conclusions from the fourth set of numerical experiments have important implications for astrophysical flows. Generally, linear instabilities are viewed as more ``reliable'' in destabilizing a flow than finite-amplitude instabilities because of concerns that the threshold for the latter may be too large. However, we will now show that the triggering threshold will be very small in protoplanetary disks.    

From our numerical experiments, we expect ZVI to be triggered when the peak vorticity on the smallest scales exceeds some critical value: $\widetilde{Ro}(k_{max}) > Ro_{crit}\sim 0.1$, where $k_{max}$ is the largest wavenumber in the flow.  This value of  the critical Rossby number assumes $\beta\equiv N/\Omega = 2$, and will be larger for smaller $\beta$. In a numerical calculation, $k_{max}$ would be $k_{res}$, but in a real fluid with viscosity, $k_{max} \sim k_{visc} \equiv 2\pi/\ell_{\nu}$, where $\ell_{\nu}$ is the viscous dissipation length.  Using Eq.~(\ref{E:Ro_scale1}), we can write the criterion for instability to be:
\begin{equation}
\widetilde{Ro}(k_{max})=\widetilde{Ro}(k_{min})\left(\frac{k_{max}}{k_{min}}\right)^{(3-a)/2} > Ro_{crit},
\end{equation}
where $k_{min}=2\pi/L$, the wavenumber for largest length scale in the flow.  Using Eq.~(\ref{E:RoMa}), we can write the instability criterion in terms of the Mach number on the largest scale:
\begin{equation}
\widetilde{Ma}(k_{min})\left(\frac{C_s}{2\Omega_0L}\right)\left(\frac{L}{\ell_{\nu}}\right)^{(3-a)/2} > Ro_{crit},
\end{equation}
where we substituted $k_{max}/k_{min} = L/\ell_{\nu}$.  For a protoplanetary disk, $C_s\approx H_0\Omega_0$.  The largest length scales with respect to turbulence will not exceed the scale height, so we also take $L\sim H_0$.  From Eq.~(\ref{E:Ma_scale3}), we can also take $\widetilde{Ma}(k_{min})\sim Ma_{rms}$.  Using Eq.~(\ref{E:reynolds}) to express the dissipation length in terms of the Reynolds number, the instability criterion becomes:
\begin{equation}
Ma_{rms} > Ro_{crit}Re^{-[(3-a)/(1+a)]} \sim Ro_{crit}\left(\frac{\ell_{mfp}}{H}\right)^{[(3-a)/(1+a)]},
\end{equation}
where the Reynolds number is $Re\equiv U\Lambda/\nu\sim H/\ell_{mfp}$, $U$ is a characteristic velocity which we take to be the sound speed, $\Lambda$ is a characteristic length which we take to be equal the scale height, $\nu\sim C_s\ell_{mfp}$ is the kinematic viscosity of an ideal gas, and $\ell_{mfp}$ is the gas mean free path.  For Kolmogorov turbulence, $a=5/3$, yielding:
\begin{equation}
Ma_{rms} > Ro_{crit}Re^{-1/2}\sim Ro_{crit}\left(\frac{\ell_{mfp}}{H}\right)^{1/2}.
\end{equation}
At 1~AU, $\ell_{mfp}/H\sim 10^{-12}$\citep{cuzzi93,chiang1997}, so $Ma_{rms} > 10^{-6}$, which corresponds to rms velocities $\sim 1$~mm/s.

Note that if the initial noise has a spectral index $a>3$, then the energy spectrum is so steep that $\widetilde{Ro}(k)$ decreases, rather than increases, with increasing $k$. In this case, the initial noise must have a much larger Mach number, of order unity, to trigger ZVI. Thus, it is extremely important when considering the stability of flows to ZVI to consider initial noise with realistic spectra. An example of an initial energy spectrum so steep that $\widetilde{Ro}(k)$ decreases with $k$ is the Gaussian initial noise used by \citet{balbus96} in arguing that Keplerian flow is stable to all purely hydrodynamic perturbations. We argued in \citetalias{MPJBHL2015} that the equilibrium flow examined in \citet{balbus96} would have been unstable to ZVI and would have produced sustained {zombie} turbulence if their calculations had included vertical gravity along with the vertically-stratified density in equilibrium with that gravity (and if the simulations had a sufficiently small grid size in the $x$ direction to resolve the critical layers). In fact, even with those changes, the calculations by \citet{balbus96} would not have exhibited ZVI because the Gaussian noise with which they initialized their simulations had $\widetilde{Ro}(k)$ decreasing with increasing $k$ and the rms velocity of the initial noise was too small.

\section{CHARACTERIZATION OF ZOMBIE TURBULENCE} \label{sec:description}

\subsection{Zombie Turbulence is Turbulence}\label{subsec:turbulence}

\begin{figure}[ht]
\epsscale{0.75}
\plotone{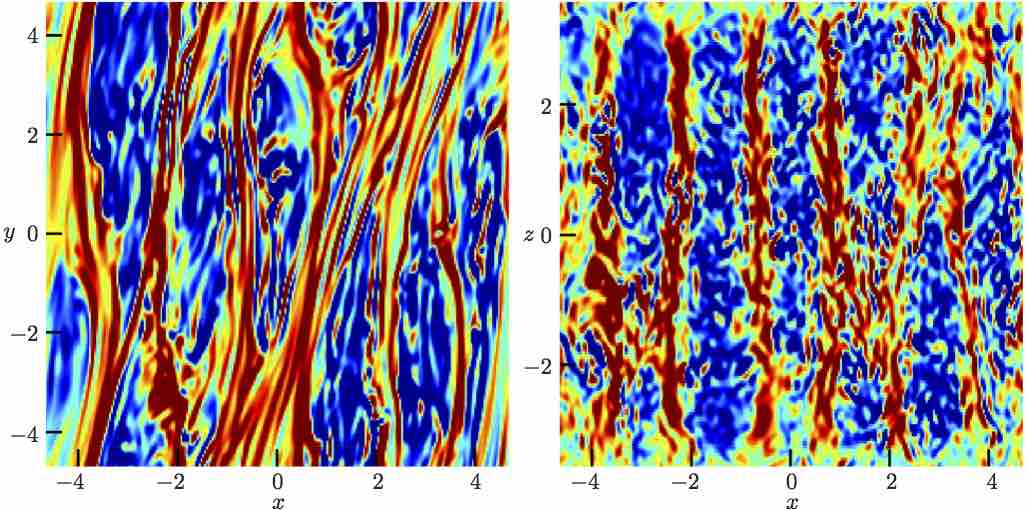}
\caption{\label{fig:zombie}
The point-wise Rossby number $Ro(x,y,z,t)$ at late-time ($\sim1370$ orbital periods) for zombie turbulence with $\beta=1.0$.   Left panel: $x$-$y$ plane. Right panel: $x$-$z$ plane at $y=0$ (but the flow is statistically invariant in $y$).  Note that the unit of length for the axes in this figure is $\Delta=H_0/3\pi$ (rather than $H_0$).  Anticyclonic vorticity is indicated by blue, while cyclonic vorticity is red; the darkest red/blue colors correspond to $Ro=\pm0.25$ while green indicates $Ro=0$.  The large-scale spatial structure is dominated by persistent anticyclonic (blue) vortices aligned in the stream-wise direction separated by cyclonic (red) vortex layers.  The cross-stream spacing between the large anticyclones is approximately $\Delta$.  Movies of these two figure are posted in the online supplementary material.}
\end{figure}

Fully-developed zombie turbulence has a well-defined large-scale spatial signature that sets it apart from other types of turbulence.  As can be seen in Fig.~\ref{fig:zombie}, the turbulent flow consists of thin cyclonic layers (red) and large anticyclonic vortices (blue) with an approximate cross-stream spacing $\Delta \equiv L_y N_0/(3 \pi \Omega_0) = L_y\beta/(3\pi)$.   The point-wise Rossby number $Ro(x,y,z,t)$ is typically $\sim$~-0.3 in the large anticyclones and $\sim$~+0.3 in the cyclonic layers.

\begin{figure}[ht]
\epsscale{0.50}
\plotone{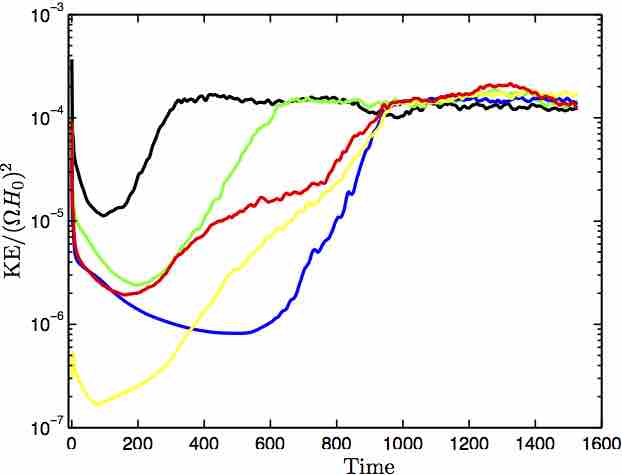}
\caption{\label{fig:attractor}
The non-Keplerian component of the kinetic energy as a function of time in units of orbital periods for five initial conditions:  Kolmogorov noise with initial rms Mach number of 0.01 (black); Kolmogorov noise with initial rms Mach number of 0.007 (green);  Kolmogorov noise with initial rms Mach number of 0.004  (blue);  noise with an energy spectral index of 1.0 and initial rms Mach number of 0.007 (red); and the run initialized with an isolated vortex with a characteristic $Ro \approx -0.5$ (yellow).  All of these initial conditions are attracted to the same statistically-steady final state, which shows the flow loses its ``memory'' of its initial conditions.  Other parameters for these simulations: $\beta=1.0$, $L_x=L_y=L_z=H_0$ resolved with $256^3$ spectral modes.}
\end{figure}

\begin{figure}[ht]
\epsscale{0.75}
\plotone{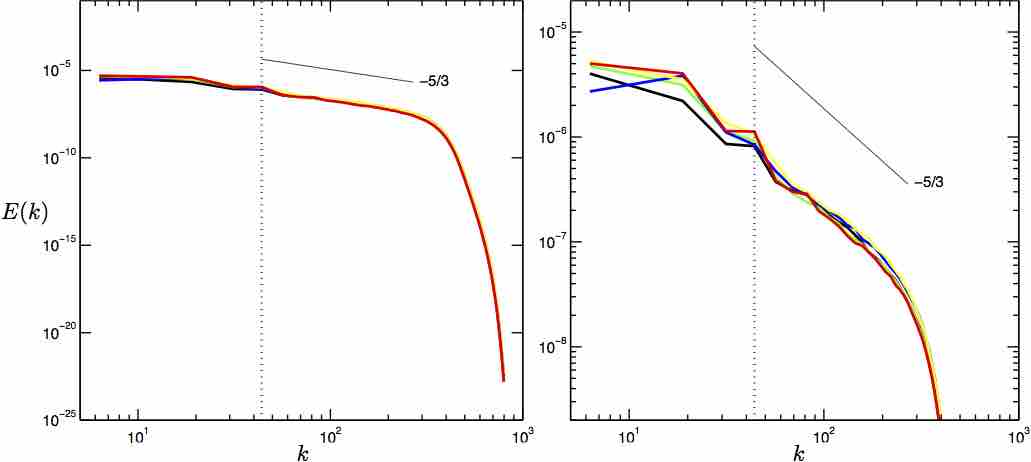}
\caption{\label{fig:spectrum}
Energy spectra $E(k)$ of the non-Keplerian component of kinetic energy of late-time zombie turbulence for the same five runs in Fig.~\ref{fig:attractor}. The right panel is a blow-up of the left panel. The figure shows the late-time flows are all attracted to the same energy spectrum, regardless of the initial conditions. Dissipation from hyperviscosity is responsible for the rapid downturn in energy at $k \gtrsim 300$. The Kolmogorov spectral index of $5/3$ is indicated by the slope of the thin solid black lines. For $80 \lesssim k \lesssim 300$, the spectra are approximately Kolmogorov, but for $k \lesssim 80$, there are departures from the $-5/3$ slope. The  peaks and valleys in this part of the spectra indicate the presence of large-scale structures in the flow. The vertical dotted line is at $k=2\pi/\Delta$, the wavenumber of the approximate cross-stream spacing of the large anticyclonic vortices and of the cyclonic layers. }
\end{figure}

On much smaller scales, zombie turbulence has properties in common with classical fully-developed homogeneous, isotropic turbulence.  Fig.~\ref{fig:attractor} shows the time-evolution of perturbation kinetic energy for five simulations that differed only in the initial conditions: three initial conditions had spectral index $a=5/3$ but different values for $E_0$, one had a spectral index $a=1$, and another was initialized with a laminar, nearly-steady coherent vortex.  All of these initial conditions triggered ZVI and developed zombie turbulence at late times.  What is interesting is that all of these initial conditions evolved toward the same final state with nearly identical values of the late-time kinetic energy within the perturbations.  Fig.~\ref{fig:spectrum} shows the energy spectra of the five flows at late time. For almost all wavenumbers $k$, the five spectra are nearly identical, supporting the conclusion that the five flows evolve toward a common turbulent attracting state. The figure shows that for mid-range and large wavenumbers ($k \gtrsim 80$), the late-time flow has spectral index of $5/3$ like classic Kolmogorov turbulence.  Perhaps this results is not surprising for the three flows that were initialized with Kolmogorov noise; however, one initial condition had a different spectral index, and another was initialized with a laminar, coherent nearly steady-state vortex (which has a decidedly non power-law energy spectrum).

The large-scale anticyclonic vortices and cyclonic vortex layers in the late-time flow in Fig.~\ref{fig:zombie} have features of both laminar and turbulent flows. The fact that the anticyclonic vortices look ``ragged'' with a great deal of small-scale variation in the magnitude of the vorticity in their interiors is in marked contrast with laminar vortices where the vorticity is smooth in their interiors.   Yet, the large anticyclones and cyclonic vortex layers in zombie turbulence are long-lived.  We define the autocorrelation function of the vertical component of the relative vorticity: $\chi(t) \equiv  \langle \omega_z (t +t')\omega_z(t')\rangle/\langle\omega_z(t')\omega_z(t')\rangle$ averaged over a sample of Lagrangian fluid elements inside a large anticyclone (where angle brackets indicate time-averaging). Defining the characteristic lifetime $T$ of a vortex by $\chi(T)=0.5$, we find that the lifetimes of anticyclones are $\sim$~30 vortex turn-around times, where the latter time is defined as $4 \pi/\omega_z$. The lifetimes of a vortex or eddy in the picture of classic turbulence is approximately one eddy turn-around time, meaning that an eddy falls apart and passes its energy onto small eddies in one turn-around \citep{tennekes1972first}. Thus, the large anticyclones in zombie turbulence have long lifetimes compared to eddies in classical turbulence. The autocorrelation time of the large-scale cyclonic  layers is of the same order as autocorrelation time of the large-scale anticyclones, so they too are long-lived. 

\subsection{Memory of the Critical Layer Eigenmodes} \label{subsec:memory}

\begin{figure}[ht]
\epsscale{0.65}
\plotone{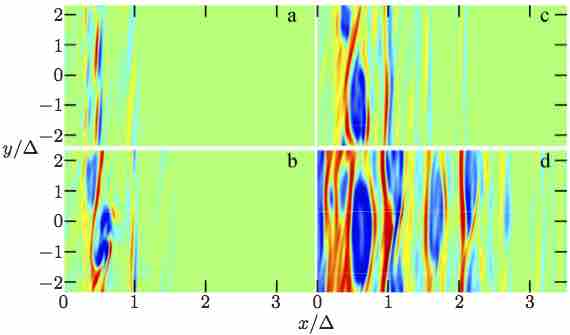}
\caption{\label{fig:zombie2}
Reproduced from \citetalias{MPJH13}. Point-wise Rossby number $Ro(x, y, z, t)$ in a $x$-$y$ plane at $z\!=\!-0.4\Delta$, for a Boussinesq flow with $\beta\!=\!2$.  The simulation was initialized with a quasi-steady state vortex in the midplane ($z\!=\!0$), so not visible at this height.  The vortex layers associated with baroclinic critical layers are clearly seen at the locations predicted by $x^*\!=\!\Delta/m$ for non-zero integer $m$.  Anticyclonic vorticity is indicated by blue, while cyclonic vorticity is red; the darkest red/blue colors correspond to $Ro=\pm0.10$ while green indicates $Ro=0$.  (a) $t = 64/N_0$, (b) $t = 256/N_0$, (c) $t = 576/N_0$, (d) $t = 2240/N_0$.}
\end{figure}

We now show that the patterns of the persistent anticyclones and cyclonic vortex layers in late-time zombie turbulence are set by the ``memory'' of the neutrally-stable critical layers, rather than initial conditions or any other properties of the flow.  In \citetalias{MPJH13}, we analyzed the linearized equations for a rotating, uniformly stratified fluid, assuming eigenmodes proportional to $\exp[i(k_y y + k_z z - s t)]$ (\ie periodic in streamwise $y$ and vertical $z$ directions).  The eigenequation obtained from this linearization (a generalization of the Rayleigh equation for the inviscid stability of shear flows \citep{drazinreid81}) is a second-order ordinary differential equation in which the coefficient of the highest-derivative term is:
\begin{equation}
A(x) = [\bar{v}_y(x) - s/k_y]\{[\bar{v}_y(x) - s/k_y]^2 - (N_0/k_y)^2\}. \label{E:eigenequation}
\end{equation}
While this was derived under the assumption of the Boussinesq approximation, we demonstrated its validity for the anelastic approximation in \citetalias{MPJBHL2015}. For $N_0> 0$, it can be shown that the flow is neutrally stable (\ie $s$ is real and eigenmodes neither grow nor decay).  Critical layers are special locations in the flow in which the coefficient of the highest-derivative term of the eigenequation is zero and the eigenmodes are singular \citep{drazinreid81}.  For $\bar{v}_y(x) = -(3/2)\Omega_0 x$, Eq.~(\ref{E:eigenequation}) vanishes for two families of critical layers; one family is located at $x^* = - sL_y/(3 m \pi \Omega_0)$, and a second family is at:
\begin{equation}
x^* = -(s\pm N_0)L_y/(3\pi m\Omega_0), \label{E:critical_layers}
\end{equation}
where $m$ is an non-zero integer.  The first family, barotropic critical layers, has already been well-studied because it also occurs in shear flows with no gravity, and/or no density stratification, and/or no rotation. It has been found that this first family of critical layers is difficult to excite, which is consistent with our own numerical experiments in \citetalias{MPJH13}.  However, the second type, baroclinic critical layers, is easily excited by perturbations; \citetalias{MPJH13} showed that when the flow is perturbed locally via a small vortex, then the excited critical layers (when observed in the Galilean frame in which the perturbing vortex is stationary) have a temporal frequency $s=0$.  The equations of motion and boundary conditions are invariant under translation in the cross-stream direction $x$ by any distance $\delta$, if there is also a Galilean shift to a frame moving in the streamwise direction $y$ with velocity $-(3/2)\Omega_0\delta$ (this is the same invariance that is exploited when one uses shearing box boundary conditions).  Due to this invariance, we are free to choose an arbitrary origin of the $x$ coordinate.  So how then do we interpret Eq.~(\ref{E:critical_layers}) with respect to the location of the critical layers?  This conundrum is solved by noting that the only physical process that breaks the invariance of the flow is the location of the perturbing vortex.  The vortex is implicitly assumed to be at the origin, and the cross-stream distance $x$ between a localized perturbation and the $s=0$ critical layer it excites is:
\begin{equation}
x^*(m) \equiv \frac{N_0}{\Omega_0}\frac{L_y}{3\pi m} = \frac{\beta L_y}{3\pi m}\equiv \frac{\Delta}{m}. \label{E:spacing}
\end{equation}
In plots of vertical velocity, it is quite easy to pinpoint the baroclinic critical layers because the vertical velocity is generally very small except in the immediate vicinity of critical layers (as in Fig.~\ref{fig:velocityeigenmode}).  Vertical vorticity is also an excellent signature because baroclinic critical layers generate vortex layers with relatively large vorticity.  The vertical component of the curl of the momentum equation Eq.~\ref{E:anelastic2} yields the vorticity equation:
\begin{equation}
\frac{\partial\omega_z}{\partial t} + (\boldsymbol{v}\cdot\boldsymbol{\nabla})\omega_z=  [(\boldsymbol{\omega}+2\Omega_0\boldsymbol{\hat{z}})\cdot\boldsymbol{\nabla}]v_z -(\omega_z+2\Omega_0)\boldsymbol{\nabla}\cdot\boldsymbol{v}.\label{E:vorticity1}
\end{equation}
With the Boussinesq approximation, $\boldsymbol{\nabla}\cdot\boldsymbol{v}=0$.  We can also separate the Keplerian shear vorticity from the perturbation vorticity: $\omega_z = \tilde{\omega}_z-(3/2)\Omega_0$.  The perturbation vorticity equation is:
\begin{equation}
\frac{\partial\tilde{\omega}_z}{\partial t} + (\boldsymbol{v}\cdot\boldsymbol{\nabla})\tilde{\omega}_z=  (\boldsymbol{\tilde{\omega}}\cdot\boldsymbol{\nabla})v_z +\tfrac{1}{2}\Omega_0\frac{\partial v_z}{\partial z}.\label{E:vorticity2}
\end{equation}
Vortex stretching and tilting create new and amplify existing vorticity, yielding cyclonic and anticyclonic vortex layers.  Fig.~\ref{fig:zombie2} (which is a reproduction of Fig.~1 from \citetalias{MPJH13}) shows $\tilde{\omega}_z$ in an $x$-$y$ plane at height $z\!=\!-0.4\Delta$ for a Boussinesq simulation with $\beta\!=\!2$.  The flow was initialized with a coherent quasi-steady 3D vortex centered on the midplane $z=0$, so the vortex core is not visible at this height.  One can readily see the vortex layers created by the baroclinic critical layers with $m=1$, $2$, and~$3$ at the locations $x^*\!=\! \Delta/m$ as predicted by Eq.~(\ref{E:spacing}). Note that the $m$ is the wavenumber in the $y$ direction, so $m$ also corresponds to the number of vortices generated within a vortex layer (see Fig.~1 in \citetalias{MPJBHL2015}). In accord with  Eq.~(\ref{E:spacing}), all of the critical layers  and vortex layers in Fig.~\ref{fig:zombie2}(a) have locations $|x^*| \le \Delta$.

Vortex layers embedded in a background shearing flow tend to be linearly stable [unstable] when the relative vertical vorticity of the layer $\tilde{\omega}_z$ has the opposite [same] sign as the vertical vorticity of the background shearing flow \citep{marcus88,marcus90a}. For example, a cyclonic vortex layer embedded in Keplerian anticyclonic shear will be stable, whereas an anticyclonic vortex layer will be linearly unstable.\footnote{This linear instability [stability] of the anticyclonic [cyclonic] vortex layers embedded in anticyclonic shear extends to layers of vortensity (or potential vorticity, as it is referred to by geophysicists \citet{pedlosky79}). \citet{marcus93} showed the instability was responsible for the roll-up of anticyclonic zonal flows on Jupiter into Great Red Spot-like large anticyclones, and \citet{lovelace1999rossby} identified the instability as  the Rossby Wave Instability when they investigated vortensity layers in accretion disks.}  The nonlinear evolution of the instability results in an anticyclonic vortex layer developing waves, which grow and break up into a streamwise series of vortices, and eventually the vortices merge together into one large anticyclone. However, a cyclonic layer (in anticyclonic shear), even when strongly perturbed and twisted, remains intact.  Figures~\ref{fig:zombie2}(b--c) show these signatures. 

Figs.~\ref{fig:zombie2}(c--d) shows another phenomenon -- the spawning of next generation zombie vortices. The vortices generated by the critical layers at $x = \Delta$ act as perturbation sources and excite new critical layers at $x = \Delta + \Delta/3$,   $x = \Delta + \Delta/2$,   and $x = 2\Delta$. The creation of a new generation of critical layers and vortex layers by the previous generation is the crux of zombie turbulence.  Clearly, the pattern of anticyclones and cyclonic layers is due to the original spacings of the critical layers as given in Eq.~(\ref{E:spacing}), and {\it not} due to the initial conditions. We remind the reader that the pattern in Fig.~\ref{fig:zombie2} was due to an initial perturbation consisting a single vortex, while the pattern in Fig.~\ref{fig:zombie} was initiated by space-filling noise.  

\subsection{Large-Scale Patterns in PPD Flows}\label{subsec:patterns_in_ppd}

Eq.~(\ref{E:spacing}) for the critical layers' positions was derived assuming the Boussinesq approximation with constant vertical gravity and constant Brunt-V\"{a}is\"{a}l\"{a} frequency.  However, in \citetalias{MPJBHL2015} and in Fig.~\ref{fig:zombie} here, we demonstrated numerically that Eq.~(\ref{E:spacing}) also successfully predicts the critical layer locations for anelastic and fully-compressible flows with constant vertical gravity and constant Brunt-V\"{a}is\"{a}l\"{a} frequency.   One can prove this formally by including density fluctuations in the linear analysis.  More importantly, it can be shown that Eq.~(\ref{E:spacing}) is valid for anelastic and fully-compressible flows with vertical gravity that is linear in $|z|$ (as in a PPD) and with flows with non-constant $N(z)$. This also can be formally proved using a WKBJ analysis, and the upshot is that it is valid to replace $N_0$ with $N(z)$ in Eq.~(\ref{E:spacing}); the locations of the critical layers therefore vary with height.  We will not present the WKBJ analysis here, nor will we now fully explore ZVI when $N(z)$ is not constant -- these are topics for future papers in our series on ZVI.  However, we do want to at least demonstrate numerically that Eq.~(\ref{E:spacing}) is indeed valid for non-constant $N(z)$.

\begin{figure}[ht]
\epsscale{0.5}
\plotone{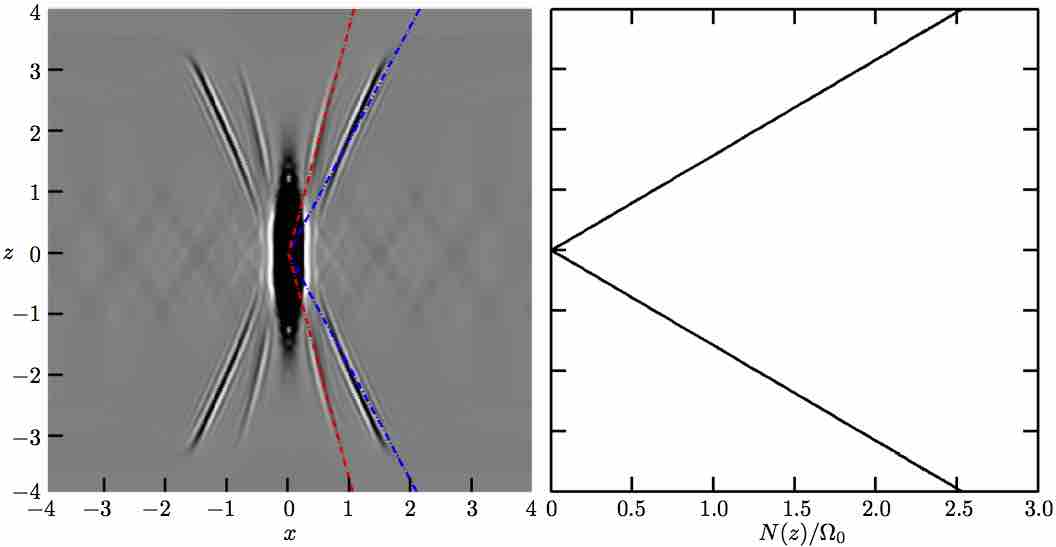}\\
\plotone{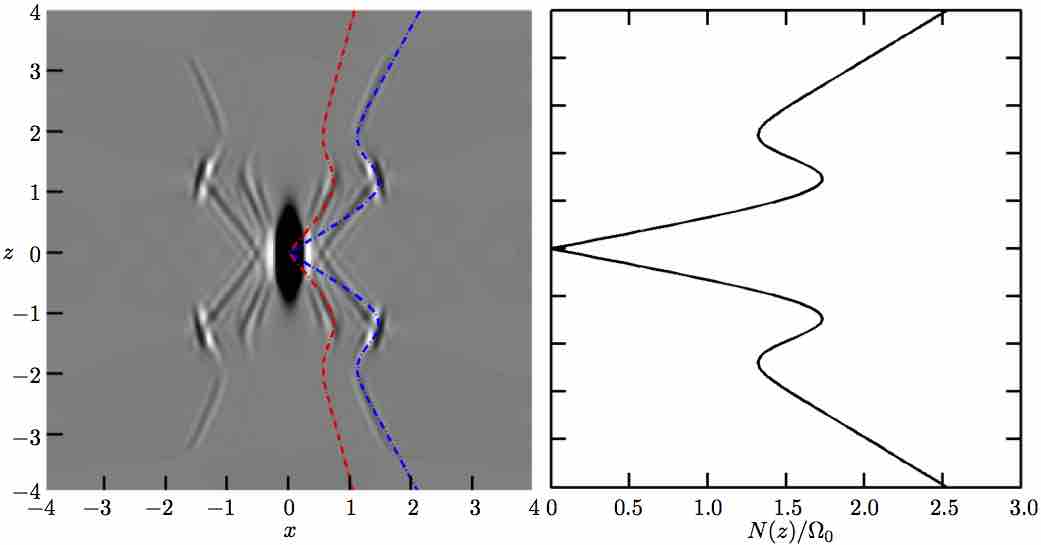}
\caption{\label{fig:zombie_ppd}
Anelastic simulations of the excitation of baroclinic critical layers by a quasi-steady state 3D perturbing anticyclone in the midplane of a PPD with linear gravity.  Units of length for $x$ and $z$ axes is $H_0$.  Integration times were short, just long enough to see vortex layers being created at the critical layers.  {\bf Left panels:} Point-wise Rossby number $Ro(x,y,z,t)$ in the $x$-$z$ plane at $y=0$.  The grey scale ranges from $-0.1$ (black) to $+0.1$ (white).  The peak Rossby number inside the vortex is $-0.3125$.  The eight thin, black-and-white nearly-straight lines extending symmetrically out of the anticyclone are vortex layers created by and at the critical layers. The dashed colored lines are the locations of the $m=1$ (blue) and $m=2$ (red) critical layers as predicted by Eq.~(\ref{E:spacing}), but with $N_0$ replaced with $N(z)$. {\bf Right panels:} Brunt-V\"{a}is\"{a}l\"{a} frequency profile $N(z)$.  {\bf Top panels:} Isothermal background, which yields $N(z)\propto|z|$. {\bf Bottom panels:} Background temperature with a cool midplane and a warm upper atmosphere; with linear gravity, this yields a Brunt-V\"{a}is\"{a}l\"{a} frequency profile $N(z)$ that has local maxima and minima.}
\end{figure}

Fig.~\ref{fig:zombie_ppd} shows anelastic simulations of the excitation of baroclinic critical layers by a quasi-steady state 3D perturbing vortex in the midplane of a PPD with linear gravity. The left panels are plots of the point-wise Rossby number $Ro(x,y,z,t)$ in the $x$-$z$ plane at $y=0$ as in Fig.~\ref{fig:zombie}, whereas the right panels show the Brunt-V\"{a}is\"{a}l\"{a} frequency $N(z)$ as functions of height.  The first row is for a uniform background temperature and gravity linear in $z$, which yields $N(z)\propto|z|$.  The second row is for a background temperature with a cool midplane and a warm upper atmosphere; with linear gravity, this yields a Brunt-V\"{a}is\"{a}l\"{a} frequency profile $N(z)$ that has local maxima and minima.  The locations of the critical layers, as predicted by Eq.~(\ref{E:spacing}) are illustrated in color: $m=\pm 1$ in blue and~$\pm 2$ in red.  The critical layers in the simulation exactly match up with the predicted locations.  \citet{barranco05} observed these patterns of the vorticity in their simulations of 3D vortices, but initially interpreted them as the St.\ Andrew cross patterns associated with propagation of internal gravity waves \citep{kundu90}.  We note that baroclinic critical layers seem to be especially excited in the vicinity of local extrema in the  Brunt-V\"{a}is\"{a}l\"{a} frequency profile $N(z)$ (bottom row); we will explore this in a future paper.

\section{SUMMARY AND FUTURE WORK}  \label{sec:discussion}

\subsection{Summary}

In the Zombie Vortex Instability, perturbations can excite baroclinic critical layers in rotating, stratified flow, which then create vortex layers that can roll-up and form anticyclones and cyclonic sheets on the large scales, and Kolmogorov-like turbulence on the smaller scales.  This is not a linear instability, but a finite-amplitude one.  The main results of this work are:

\begin{enumerate}
\itemsep0em
\item For initial perturbations with a power law energy spectrum, the threshold for triggering ZVI is set by the magnitude of the vorticity (or Rossby number) or the enstrophy on the small scales, and not on the magnitude of the velocity (or Mach number) or energy.  For $\beta\equiv N/\Omega=2$, we find that the minimum Rossby number needed for instability is $Ro_{crit}\sim0.2$ on the smallest scales.  

\item While the threshold for ZVI is set by vorticity or Rossby number, it is nonetheless useful to infer a criterion on the Mach number.  We find that the critical Mach number scales with the inverse square root of the Reynolds number: $Ma_{crit}\sim Ro_{crit}Re^{-1/2}$.  In protoplanetary disks, this is $Ma_{crit}\sim10^{-6}$.

\item On the small scales, zombie turbulence has no memory of the initial conditions and has a Kolmogorov-like energy spectrum.

\item On the large scales, zombie turbulence is characterized by anticyclones and cyclonic sheets with typical Rossby number $\sim0.3$.  The spacing of the cyclonic sheets and anticyclones appears to have a ``memory'' of the spacing of the baroclinic critical layers that gave rise to the vortex layers.  The wavenumber for this spacing is evident in the energy spectrum before the turnover to the inertial range of the Kolmogorov turbulence. 

\item While our earlier work was in the limit of uniform stratification, we have demonstrated that our formula for the critical layer spacing is also valid for non-uniform Brunt-V\"{a}is\"{a}l\"{a} frequency profiles that may be found in protoplanetary disks.

\end{enumerate}

\subsection{Future Work}

In \S\ref{subsec:patterns_in_ppd}, we demonstrated that ZVI operates in flows with non-uniform vertical gravity and non-uniform stratification, as would be found in protoplanetary disks.  In a future paper, we will more fully investigate spatially varying Brunt-V\"{a}is\"{a}l\"{a} frequency $N(z)$, with a focus on two issues: the effects of local extrema in $N(z)$, and determining how close zombie turbulence can penetrate into the unstratified midplane.

In \citetalias{MPJH13}, \citetalias{MPJBHL2015}, and this work, we worked in the adiabatic limit.  Cooling and radiative diffusion will suppress temperature anomalies in the baroclinic critical layers, which may effectively make them barotropic , halting ZVI.  A future paper will address relaxing adiabaticity and simulating ZVI with both optically thin cooling and optically thick radiative diffusion, with a focus on determining the minimum cooling times that can sustain zombie turbulence.  Optically thick cooling works on all size scales, whereas radiative diffusion operates most effectively on small size scales, especially the scale of critical layers.  Two other purely hydrodynamic instabilities, vertical shear instability (VSI) and convective overstability (ConO), also operate in PPDs, but these require very short cooling times.  A potential area of study could be the interplay of these instabilities in regions of parameter space where they may overlap.

Of high priority is investigating the role of ZVI in star and planet formation.  We plan to undertake simulations with a new spectral fully-compressible code in order to compute angular momentum transport via compressible modes excited by zombie turbulence.  With a new dust transport code, we will also pursue the dust trapping properties of coherent zombie vortices.   Of special interest will be to look at the competition or interplay of ZVI with the streaming instability.  Finally, one might envision investigating ZVI with MRI, in global simulations, or in laboratory flows.

Magnetohydrodynamic processes (MRI, MHD-induced winds) and purely hydrodynamic instabilities (ConO, VSI, ZVI) may leave their own unique signatures on dust emission.  In the age of ALMA (and with JWST in the very near future), we can expect direct observations of these dust patterns which might allow us to discriminate among the various mechanisms that drive turbulence within protoplanetary disks \citep{ruge2016,alma2015}

\acknowledgments

PSM is supported by NSF grants AST-1009907 and AST-1510703 and by NASA PATM grants NNX10AB93G and NNX13AG56G.   JAB is supported by NSF grants AST-1010052 and AST-1510708.  Support for computational work comes NSF XSEDE (NSF OCI-1053575) and NASA-HEC.  PSM \& JAB would like to thank the Kavli Institute for Theoretical Physics (KITP) for hosting us while drafting this paper (NSF grant PHY-1125915).


\end{document}